\documentclass[intlimits,twoside,a4paper]{article}

\usepackage[cp1251]{inputenc}
\usepackage[eqsecnum]{cmpj3}
\usepackage{bm}

\issue{2022}{25}{1}{13701}
\doinumber{10.5488/CMP.25.13701}
\title[Electronic structure of short-period ZnSe/ZnTe superlattices]%
{Electronic structure of short-period ZnSe/ZnTe superlattices based on DFT calculations}
\author[M. Caid \emph{et al.}]{M.~Caid\orcid{0000-0001-9773-8877}\refaddr{label1,label2}, 
        Y.~Rached\refaddr{label3},  D.~Rached\orcid{0000-0003-4686-5686}\refaddr{label2}, O.~Cheref\orcid{0000-0002-3672-1083}\refaddr{label2}\thanks{Corresponding author: \email{oualidcheref@gmail.com}.},  H.~Rached\orcid{0000-0003-3867-576X}\refaddr{label2,label4},  S.~Benalia\orcid{0000-0003-0326-8056}\refaddr{label2,label3},  M.~Merabet\refaddr{label2,label3}} 
\addresses{
\addr{label1} \'{E}cole Normale Sup\'{e}rieure de Bou Sa\^{a}da, D\'{e}partement des Sciences Exactes, Bou Sa\^{a}da (28001), M'sila, Alg\'{e}rie
\addr{label2} Laboratoire des Mat\'{e}riaux Magn\'{e}tiques, Facult\'{e} des Sciences Exactes, Universit\'{e} Djillali Liab\`{e}s de Sidi Bel-Abb\`{e}s, Sidi Bel-Abb\`{e}s (22000), Alg\'{e}rie
\addr{label3} D\'{e}partement des Sciences de la Mati\`{e}re, Facult\'{e} des Sciences et de la Technologie, Universit\'{e} Ahmed Ben Yahia El-Wancharisi Tissemsilt, Tissemsilt (38000), Alg\'{e}rie
\addr{label4} Hassiba Benbouali university of Chlef, Faculty of Exact Sciences and Informatic, Department of physics, Chlef~(02000), Alg\'{e}rie
}


\date{Received July 17, 2021, in final form November 24, 2021}

\begin{document}
\maketitle

\begin{abstract}
In the present study we discuss the effect of variation in the number of monolayers $n$ on the electronic and optical properties of superlattices (SLs) (ZnSe)$_n$/(ZnTe)$_n$. The total energies were calculated by the full-potential linear muffin-tin orbital (FP-LMTO) method, and the exchange-correlation energy was applied in the local density approximation (LDA). First, the calculations show a decrease in the derivative of bulk modulus and electronic bandgap with an increase in the number of monolayers $n$. Second, the radiation energies up to $15$~eV, the dielectric function $\varepsilon(\omega$), the refractive index $n(\omega)$, and the reflectivity $R(\omega)$ are studied. These calculations may be beneficial to understand the properties of short-period superlattices (ZnSe)$_n$/(ZnTe)$_n$.
%
%
\printkeywords
\end{abstract}

\section{Introduction}

Superlattices play an important role in technological development due to their electronic characteristics and photonic diversity. Among these, superlattices II-VI are widely theoretically and experimentally studied. For example, (ZnSe)$_n$/(ZnTe)$_n$ superlattices (SLs) are used due to their belonging to II-VI semiconductors compounds, which have direct band gaps in the visible region of the spectrum~\cite{Raj88}. Moreover, ZnSe/ZnTe superlattices  offer a considerable promise as a means of producing tunable light emitters by varying the thickness of the constituent material layer~\cite{Raja88}. These SLs may be used in manufacturing blue light-emitting diodes (LEDs) and short-wavelength semiconductor lasers~\cite{Kob86,Koba86,Kobay86,Mil87}. The successful growth of these superlattices by molecular-beam epitaxy (MBE) and preliminary characterization studies have already been reported~\cite{Kob87}. Photoluminescence from these SLs was studied by Kobayashi et al.~\cite{Koba86} and Kuwabara et al.~\cite{Kuw86}. In 2017, Zhitov et al.~\cite{Zhi17} fabricated and investigated the type-II ZnSe/ZnTe SLs for photodetector application. From the Laboratoire des
Mat\'{e}riaux Magn\'{e}tiques (LMM), we investigated the structural, electronic, and optical properties of II-VI and III-V SLs by using the full-potential linear muffin-tin orbital (FP-LMTO) method, among the BeTe/ZnSe~\cite{Cai16}, InAs/GaSb~\cite{Cai19}, ZnTe/MnTe~\cite{Adi20}, and AlSb/GaSb~\cite{Cai20}. We found that the properties of SLs are intensely layers dependent and SLs show a potential for technological applications.

The advantages of the properties of bulk ZnX (X=Se, Te) compounds and their (ZnSe)$_n$/(ZnTe)$_n$ superlattices for eventual technological applications are obvious, and the studies of structural, electronic and optical properties of short-period (ZnSe)$_n$/(ZnTe)$_n$ ($n \leqslant 3$) are necessary. In this work, we probe the properties of the (ZnSe)$_n$/(ZnTe)$_n$ ($n$ is number of monolayers; $n = 1, 2$ and $3$) SLs, using the FP-LMTO method within the density functional theory (DFT). The arrangement of this work is as follows. In section~2, we reported the utilized method with the disclosure of calculations. Results and discussions of the structural, electronic, and optical properties of (ZnSe)$_n$/(ZnTe)$_n$~SLs are presented in section~3. Finally, conclusions and remarks are given in section~4.
\section{Calculation methodology}

Calculations of our work are performed by using the FP-LMTO method implemented in the new version LmtART computer code~\cite{Sav92,Sav96}. The local-density approximation (LDA) was used to describe the exchange-correlation potential~\cite{Per92}. These calculations are based on the DFT, which is a universal quantum mechanical approach for many-body problems. In this theory, interacting electrons and nuclei are represented in a system of one-electron equations called Kohn--Sham equations~\cite{Hoh64,Koh65}.

\begin{table}[htb]
	\caption{Parameters used in the calculations: the number of plane wave (NPLW), energy cutoff (in Rydbergs) and the muffin-tin radius (RMT, in atomic units).}
	\label{Table 1}
	
	\begin{center}
		\begin{tabular}{||c |c |c| c c c ||} 
			\hline
		Compound & NPLW & $E_{\rm{cut}}$ (Ry) &  & MTS (a.u.) &  \\ [0.5ex] 
			\hline
			& & &Zn  & Se & Te \\
			\hline
			ZnSe & 5064 & 99.5033 & 2.221 & 2.406 & - \\  
			\hline
			ZnTe & 5064 & 89.1336 & 2.249 & - & 2.640 \\
			\hline
			
			(ZnSe)${}_{1}$/(ZnTe)${}_{1}$ & 16242 & 127.0687 & 2.248 & 2.529 & 2.529 \\
			\hline
			(ZnSe)${}_{2}$/(ZnTe)${}_{2}$ & 32458 & 126.7653 & 2.237 & 2.487 & 2.558 \\ 
			\hline 
			(ZnSe)${}_{3}$/(ZnTe)${}_{3}$ & 48690 & 126.5476 & 2.234 & 2.490 & 2.585 \\ 
			\hline 
		\end{tabular}
		
	\end{center}
\end{table}

The charge density and potential inside the MTSs are represented by spherical harmonics up $l_{max} = 6$ to achieve the energy eigenvalues convergence. The tetrahedron method~\cite{Blo94} is used for $k$ integration over the Brillouin zone (BZ), and is set up differently following the case. For SL~($n$, $n$), meshes of (6, 6, 6), (8,~8,~8) and (10, 10, 10) are utilized for SL~(1-1), SL~(2-2) and SL~(3-3), respectively. The self-consistent calculations are considered to be converged within $10^{-6}$~Ry for the total energy. The values of the sphere radii (MTS) and the number of plane waves (NPLW) used in the present calculations are listed in table~\ref{Table 1}. 

\section{Results and discussion}
\subsection{Structural properties}

As previously described (ZnSe)$_n$/(ZnTe)$_n$ SLs crystallizes in a tetragonal point group $D_{2d}$ symmetry, which $n$ is the number of monolayers ZnX~(X=Se, Te). There is no experimental and theoretical parameter to compare with our results obtained. The total energies as a function of unit cell volumes are fitted by  the Birch--Murnaghan equation of state (EOS)~\cite{Mur47} to determine the lattice constants $a_{0}$, bulk modulus~$B_{0}$, and its pressure derivatives $B_0^{\prime}$. Thus, first we calculate the lattice parameter of the ZnX~(X=Se, Te) in the space group F-43m (No.~216). Then, we study the (ZnSe)$_n$/(ZnTe)$_n$ SLs, which we choose to be the [100] as the axis of growth. Since the lattice mismatch for ZnSe/ZnTe SLs is 5.69\% between its two constituent bulk semiconductors, the lattice mismatch problem does not arise. In table~\ref{Table 2} and table~\ref{Table 3}, we summarize the calculated lattice constants and the bulk module and its pressure derivative of binaries ZnX~(X=Se, Te) and (ZnSe)$_n$/(ZnTe)$_n$ SLs together with the available experimental and theoretical data.

\begin{table}[htb]
\caption{The calculated lattice parameter $a$~(\AA), bulk modulus $B_{0}$~(GPa) and its pressure derivative $B_0^{\prime}$ for zinc blende ZnSe and ZnTe compounds at equilibrium volume compared to the available theoretical and experimental data.}
\label{Table 2}

\begin{center}
\renewcommand{\tabcolsep}{0.30pc}
\begin{tabular}{||c||c c c| c c c|c c c||} 
 \hline
 \small Compounds & &\small $a$~(\AA)& & &\small $B_{0}$~(GPa) & & &\small $B_0^{\prime}$ &   \\ [0.ex] 
 \hline
     &\small Present &\small Exp. &\small Theor. &\small  Present &\small Exp. &\small Theor.&\small Present &\small Exp. &\small Theor. \\
     
\hline
 \small ZnSe&\small 5.65 &\small 5.667\textbf{${}^{a}$} &\small 5.624\textbf{${}^{b}$}  &\small 59.68  &\small 64.7\textbf{${}^{a}$} &\small 71.82\textbf{${}^{b}$} &\small 3.96 &\small 4.77\textbf{${}^{a}$} &\small 4.88\textbf{${}^{b}$} \\
 
 &  &  &\small 5.618\textbf{${}^{c}$}  &   &  &\small 67.6\textbf{${}^{c}$} &  &  &\small 4.67\textbf{${}^{c}$} \\
 
  &  &  &\small 5.666\textbf{${}^{d}$}  &   &  &\small 67.32\textbf{${}^{d}$} &  &  &\small 4.599\textbf{${}^{e}$} \\
  
  &  &  &\small 5.578\textbf{${}^{e}$}  &   &  &\small 71.84\textbf{${}^{e}$} &  &  &\small 4.57\textbf{${}^{f}$} \\
  
  &  &  &\small 5.611\textbf{${}^{f}$}  &   &  &\small 75.20\textbf{${}^{f}$} &  &  &\small 3.96\textbf{${}^{g}$} \\
  
  &  &  &\small 5.65\textbf{${}^{g}$}  &   &  &\small 59.68\textbf{${}^{g}$} &  &  &  \\
 \hline
 \small ZnTe&\small 5.97 &\small 6.00\textbf{${}^{h}$}  &\small 5.98\textbf{${}^{i}$} &\small  52.21 &\small 51\textbf{${}^{h}$} &\small 54.03\textbf{${}^{i}$} &\small 4.86 &\small 4.70\textbf{${}^{h}$} &\small 4.98\textbf{${}^{i}$} \\
 
 &  &  &\small 6.00\textbf{${}^{j}$} &   &  &\small 52.21\textbf{${}^{k}$} &  &  &\small 4.86\textbf{${}^{k}$} \\
 
 &  &  &\small 5.97\textbf{${}^{k}$} &   &  &\small 51.20\textbf{${}^{l}$} &  &  &\small 4.70\textbf{${}^{m}$} \\
 
 &  &  &  &   &  &\small 55.90\textbf{${}^{n}$} &  &  &\small 4.90\textbf{${}^{o}$} \\
 \hline
 \multicolumn{10}{|c|}{\small ${}^{a}$ Reference~\cite{Lee70},${}^{\ }$ ${}^{b}$~\cite{Khe06}, ${}^{c}$~\cite{Cas98}, ${}^{d}$~\cite{Gan03}, ${}^{e}$~\cite{Oko03}, ${}^{f}$~\cite{Rab03}, ${}^{g}$~\cite{Caid16}, ${}^{h}$~\cite{Ada05}, ${}^{i}$~\cite{Hak12},}\\
 \multicolumn{10}{|c|}{\small ${}^{j}$~\cite{Roz09}, ${}^{k}$~\cite{Bou14}, ${}^{l}$~\cite{Kan03}, ${}^{m}$~\cite{Fra03}, ${}^{n}$~\cite{Wei99}, ${}^{o}$~\cite{Mer05}.}\\
 \hline
\end{tabular}
\end{center}
\end{table}

As can be seen, the obtained lattice constants for ZnX~(X=Se, Te) are in good agreement with the experimental data, where the difference is 0.3\%  for ZnSe and 0.65\% for ZnTe, which ensures the reliability of the present first-principles computations. It was found that the calculated compressibiity modulus and its pressure derivative for ZnSe are respectively 7.76\% and 16.98\% lower than the experimental values, while for ZnTe they are respectively 2.37\% and 3.40\% higher than the experimental values. The lattice parameter of (ZnSe)$_n$/(ZnTe)$_n$ SLs ($n=1, 2$ and $3$) shows that $a_{1-1}$ is equivalent to $[a_{0}$(ZnSe)$+a_{0}$(ZnTe)$]/2$, $a_{2-2}$ is about twice  larger  than $a_{1-1}$ and $a_{3-3}$ is threefold larger than $a_{1-1}$. The lattice constant $a_{0}$ of (ZnSe)$_n$/(ZnTe)$_n$ SLs relies on $n$ (the number of monolayers). In table~\ref{Table 3}, the lattice constant of (ZnSe)$_n$/(ZnTe)$_n$ SLs increases with the number of monolayers $n$ increasing.

\begin{table}[htb]
\caption{The calculated equilibrium constant $a$~(\AA), bulk modulus $B_{0}$~(GPa) and its pressure derivative~$B_0^{\prime}$ for superlattices (ZnSe)$_n$/(ZnTe)$_n$.}
\label{Table 3}

\begin{center}
\begin{tabular}{||c| c| c| c||} 
 \hline
 Superlattices & $a$ (\AA) & $B_{0}$ (GPa) & $B_0^{\prime}$ \\ 
 \hline 
(ZnSe)${}_{1}$/(ZnTe)${}_{1}$ & 5.84 & 58.51 & 4.74 \\ \hline 
(ZnSe)${}_{2}$/(ZnTe)${}_{2}$ & 11.69 & 58.80 & 4.26 \\ \hline 
(ZnSe)${}_{3}$/(ZnTe)${}_{3}$ & 17.55 & 58.50 & 3.74 \\ 
 \hline 
\end{tabular}

\end{center}
\end{table}

\subsection{Electronic properties}

In this section, the electronic properties of ZnX (X=Se, Te) and their SLs~($n\text{-} n$)  ($n= 1, 2$ and $3$) at this equilibrium lattice constants are described. In table~\ref{Table 4} and table~\ref{Table 5}, we listed the obtained direct band-gaps, at the high-symmetry point $\Gamma$ in the BZ, for the systems investigated and compared the results with the available theoretical and experimental data.

\begin{figure}[!htb]
	\centerline{\includegraphics[scale=0.80]{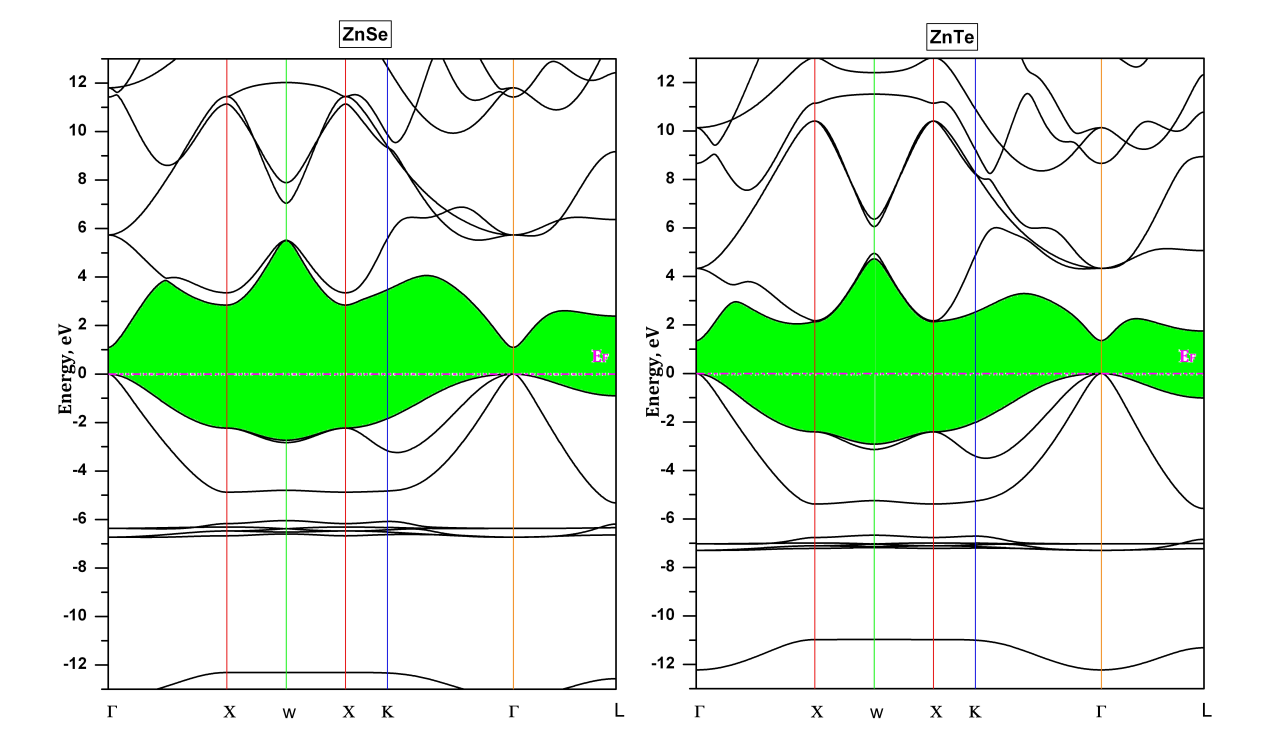}}
	\caption{(Colour online) Band structure along the symmetry lines of the Brillouin zone  for ZnSe and ZnTe compounds.} \label{figure1}
\end{figure}

\begin{figure}[htb]
	\centerline{\includegraphics[scale=0.85]{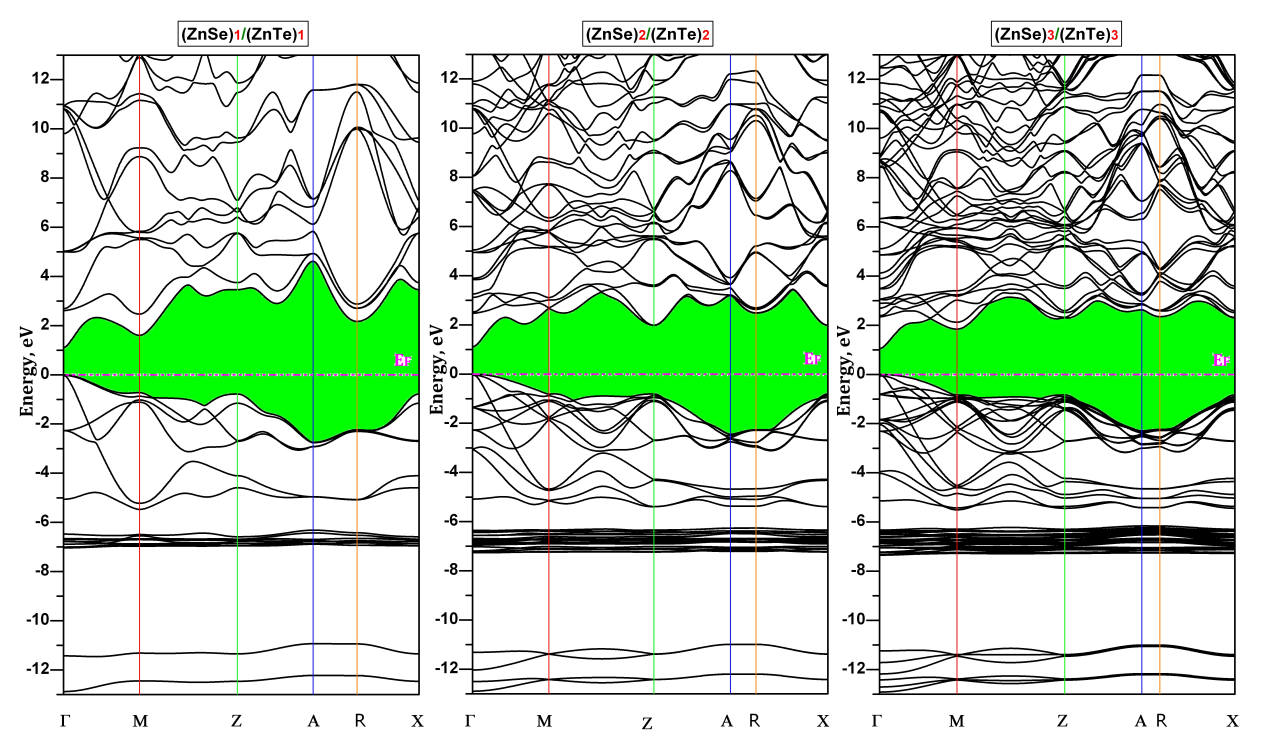}}
	\caption{(Colour online) Band structure along the symmetry lines of the Brillouin zone for (ZnSe)$_n$/(ZnTe)$_n$ superlattices.} \label{figure2}
\end{figure}

The bandgaps $E_{g}$~(eV) were calculated for the ZnX (X=Se, Te), and are in perfect agreement with the obtainable theoretical results. We can also see that, due to the approximations used in our work, the $E_{g}$ values are underestimated compared to the experimental value. Figure~\ref{figure1} shows the calculations of band structures (BS) for ZnX~(X=Se, Te). The difference in the calculated values of the band gaps compared to the experimental values is due to the well-known fact that in the electronic band structure the calculations within DFT, LDA underestimate the bandgap semiconductors. The calculated $E_{g}$ of SLs~($n\text{-} n$) as a function of the layer thickness $n$ along the higher symmetry directions of BZ is presented in figure~\ref{figure2}. According to the calculations, we find a direct bandgap ($E_{g}$) ($\Gamma$v-$\Gamma$c) of SLs~($n\text{-} n$), which is of interest for optoelectronic devices.

\begin{table}[htb]
\caption{The calculated energy band gaps for ZnSe and ZnTe in (BZ) structure.}
\label{Table 4}

\begin{center}
\begin{tabular}{||c| c c c||} 
 \hline
 Compounds &  & $E_{g}$ (eV) &  \\ 
 \hline 
 & Present & Exp. & Theor. \\ 
 \hline 
ZnSe & 1.096 & 2.82{${}^{a}$} & 1.31{${}^{b}$}\\
 &  &  & 1.83{${}^{c}$}\\
 &  &  & 1.863{${}^{d}$}\\
 &  &  & 1.0963{${}^{e}$}\\
 \hline
 ZnTe & 1.340 & 2.38{${}^{f}$} & 1.00{${}^{g}$}\\
 &  &  & 1.24{${}^{h}$}\\
 &  &  & 1.33{${}^{i}$}\\
 &  &  & 1.40{${}^{g}$}\\
 &  &  & 1.16{${}^{j}$}\\
 &  &  & 1.32{${}^{k}$}\\
 &  &  & 1.34{${}^{l}$}\\
\hline
\multicolumn{4}{|c|}{${}^{a}$ Reference~\cite{Ven79}, ${}^{b}$ \cite{Khe06}, ${}^{c}$ \cite{Wan81}, ${}^{d}$ \cite{Elh07}, ${}^{e}$ \cite{Caid16}, ${}^{f}$ \cite{Lid92},} \\
\multicolumn{4}{|c|}{${}^{g}$ \cite{Vog96}, ${}^{h}$ \cite{Roz09}, ${}^{i}$ \cite{Zak94}, ${}^{j}$ \cite{Ak09}, ${}^{k}$ \cite{Chen96}, ${}^{l}$ \cite{Bou14}}\\
\hline
\end{tabular}
 
\end{center}

\end{table}

It is important to note that the results found are automatically small, because we should not forget that the LDA method greatly underestimates the $E_{g}$ in semiconductors. 

From table~\ref{Table 5}, we can see that the $E_{g}$ decreases with an increase of the number of monolayers $n$. We note that the difference in the $E_{g}$ of the SLs~($n\text{-} n$) is usually attributed to the charge transfer, reflected in terms of the electronegativity difference of the mixing cations and anions.

\begin{figure}[htb]
\centerline{\includegraphics{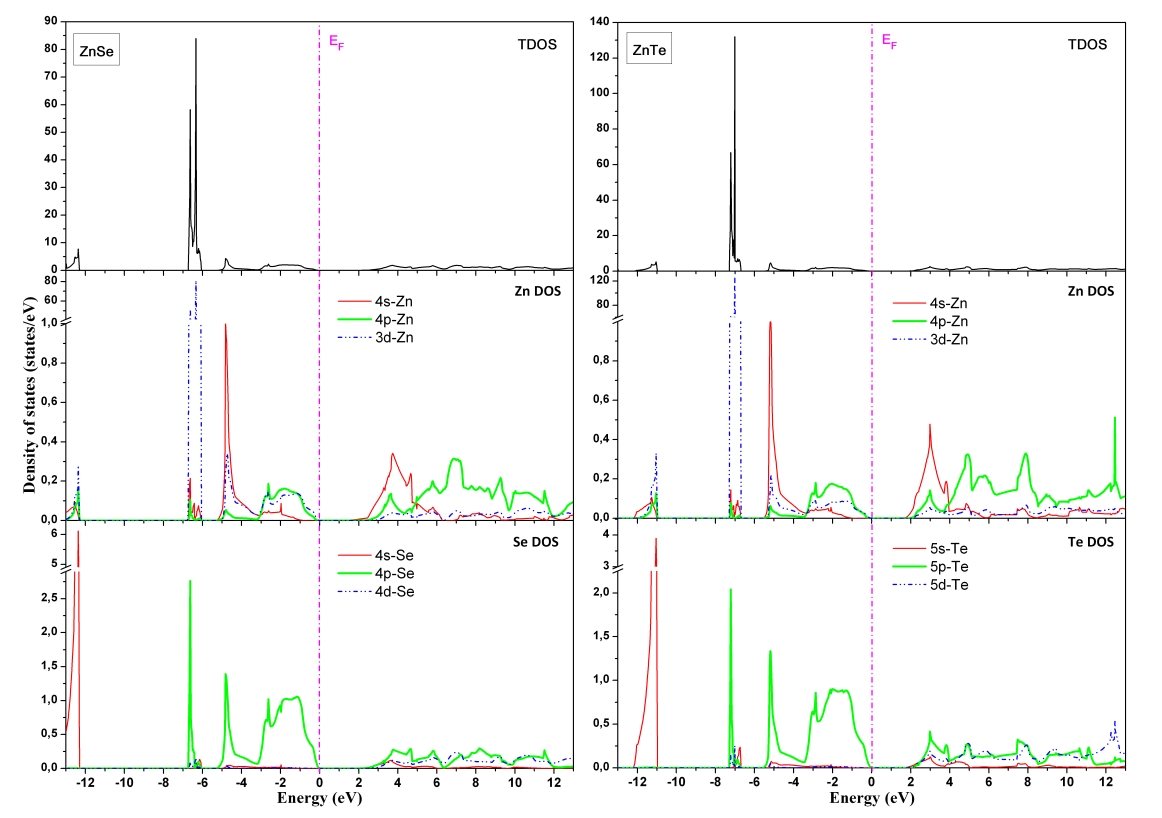}}
\caption{(Colour online) Total and partial density of states for ZnSe and ZnTe compounds.} \label{figure3}
\end{figure}

We also calculated the total and the partial densities of states (TDOS and PDOS) of ZnX (X=Se,~Te) and SLs~($n\text{-} n$), as displayed in figure~\ref{figure3} and figure~\ref{figure4}, respectively. Inspection of figure~\ref{figure3} shows that there are four distinct structures in the density of electronic states separated by gaps. The first structure (from~$-13$ to $-12.2$~eV for ZnSe) and (from $-13$ to $-11$~eV for ZnTe), respectively, originates from $4s$-Se states with a few contributions of zinc states for ZnSe, and $5s$-Te states for ZnTe with a few contributions of zinc states. The second structure (from $-6.8$ to $-6.0$~eV for ZnSe) and (from $-7.2$ to $-6.8$~eV for ZnTe), generally, comes from $3d$-Zn states with a weak contribution of $4s/4p$-Zn, $4s/4p$-Se and $5s/5p$-Te states. The third structure (from $-5.5$ to $-0.0014$~eV for ZnSe) and (from $-5.2$ to $-0.0009$~eV for ZnTe) below the Fermi level ($E_{\rm F}$) is due to the $4s/4p/3d$-Zn and $4p$-Se for ZnSe and is due to the $4s/4p/3d$-Zn and $5p$-Te for ZnTe. 

In the conduction bands, the dominant contribution is due to the states of the elements Zn and Se for ZnSe (Zn and Te for ZnTe). Examination of figure~\ref{figure4} shows that the topology of the SLs~($n\text{-} n$) densities of states is the same as that of the binary compound. The lower occupied bands, located between $-13$ and $10.9$~eV, are primarily formed by the $5s$-Te, with small contributions by the $4s/4p/3d$-Zn states. The states in the middle region between $-7.2$ and $-6.3$~eV occur mainly because of the $3d$-Zn states, with small contributions due to the other $5s/5p/5d$-Te states and the $4p$-Zn states. The states in the last region between $-5.3$~eV and the Fermi level are dominated by the $4s/3d$-Zn states. In the conduction band, we observe the dominance of $s/p/d$ states of all atoms, and we can also see the forceful hybridizations between the $4p$-Se and $4d$-Se states.

\begin{table}[htb]
\caption{The calculated energy band gaps for superlattices (ZnSe)$_n$/(ZnTe)$_n$.}
\label{Table 5}

\begin{center}
\begin{tabular}{||c| c| c||} 
 \hline
 Superlattices & $E_{g}$~(eV) & Nature \\ 
 \hline 
(ZnSe)${}_{1}$/(ZnTe)${}_{1}$ & 1.116 & Gap direct  \\ 
\hline 
(ZnSe)${}_{2}$/(ZnTe)${}_{2}$ & 1.113 & Gap direct  \\
\hline 
(ZnSe)${}_{3}$/(ZnTe)${}_{3}$ & 1.059 & Gap direct \\ 

 \hline 
\end{tabular}

\end{center}
\end{table}

\subsection{Optical properties}

\begin{figure}[!htb]
\centerline{\includegraphics{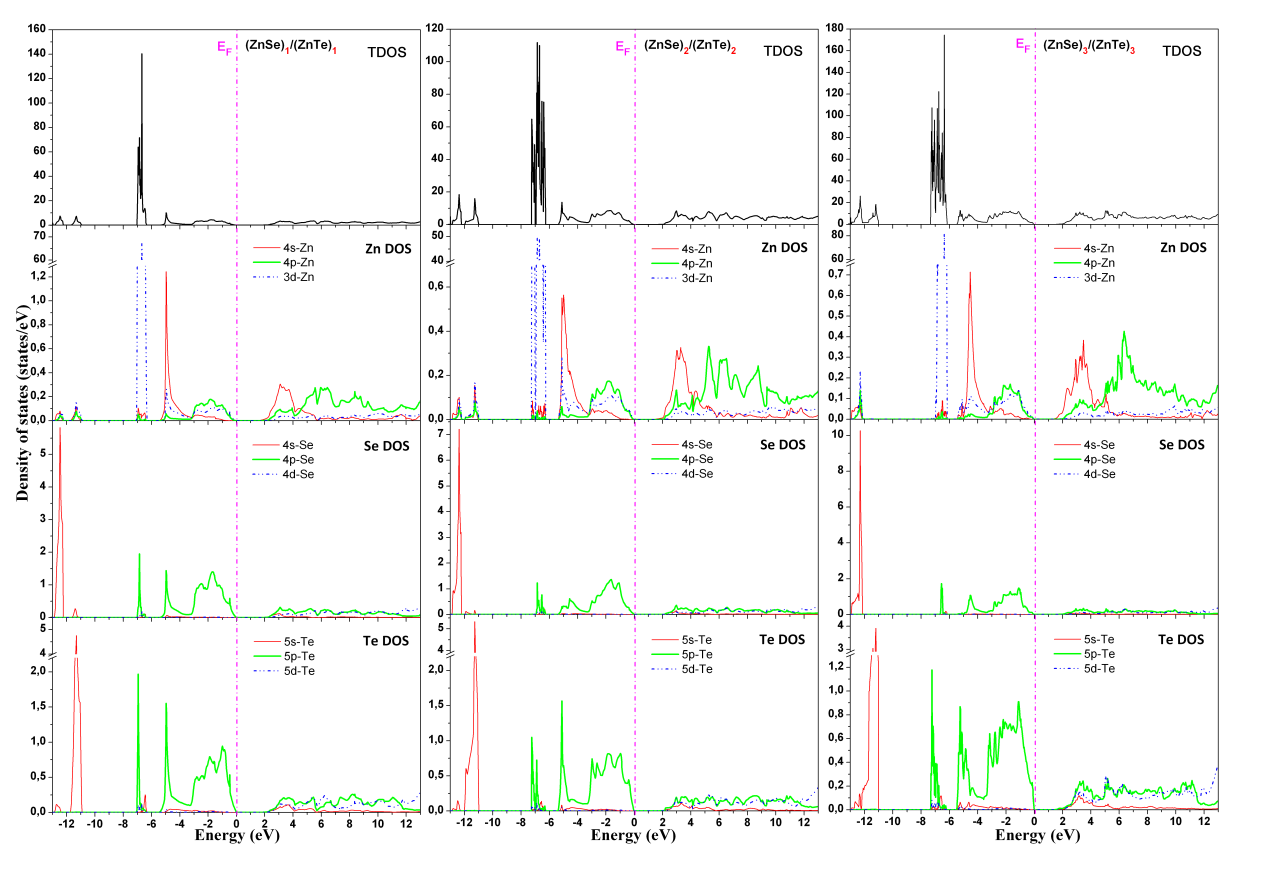}}
\caption{(Colour online) Total and partial density of states for (ZnSe)$_n$/(ZnTe)$_n$ superlattices.} \label{figure4}
\end{figure}

\begin{figure}[!htb]
	\centerline{\includegraphics[scale=0.45]{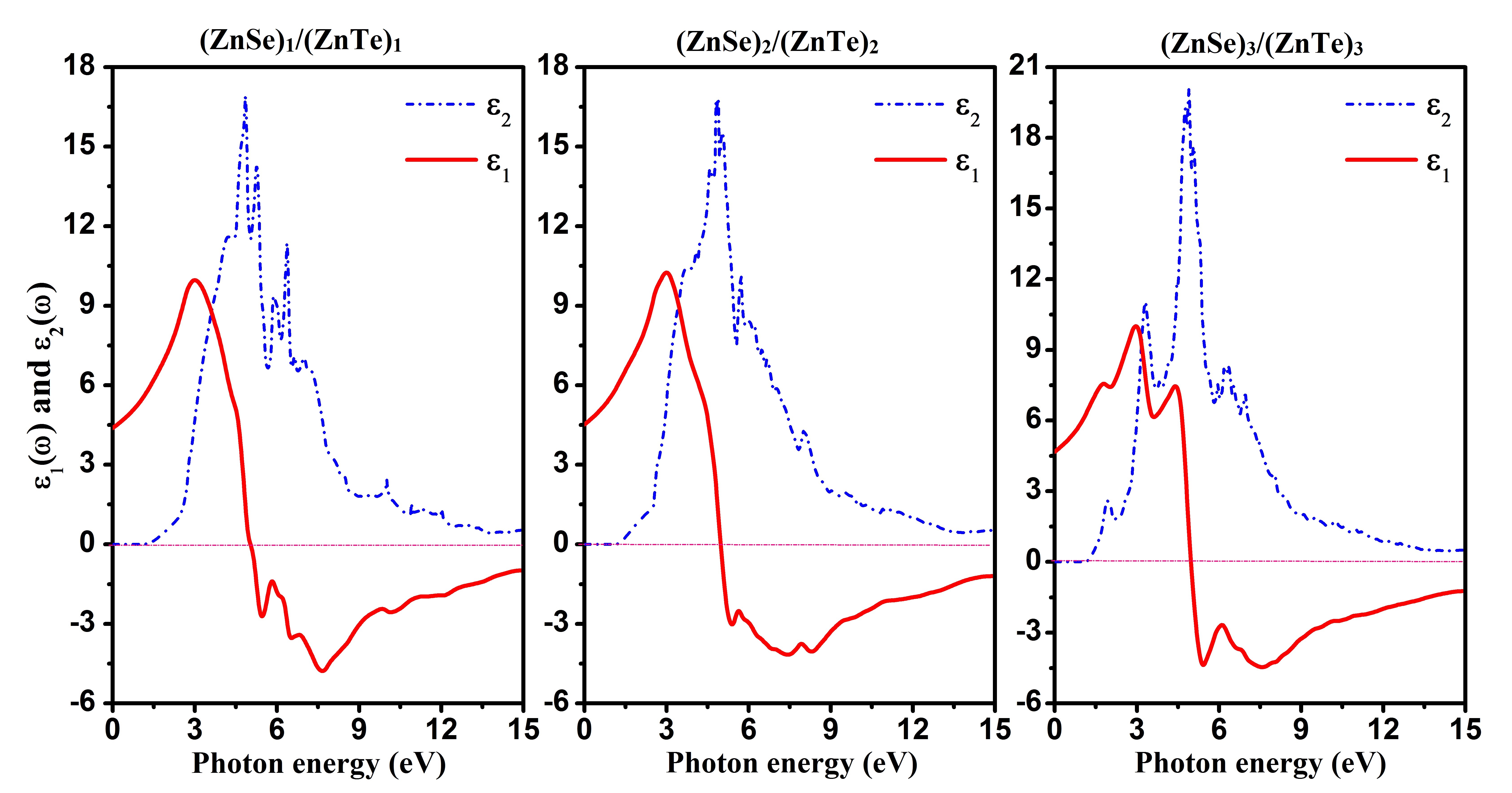}}
	\caption{(Colour online) The calculated dielectric functions (real and imaginary) for (ZnSe)$_n$/(ZnTe)$_n$ superlattices.} \label{figure5}
\end{figure}

It is very important to reach a good comprehension of the electronic behavior and deeply understand the electronic properties illustrated in figure~\ref{figure2} and figure~\ref{figure4} for our superlattices. A detailed description of the optical investigation can be found in reference~\cite{Amb06}. In this section, the potential optical properties such as real~$\varepsilon_{1}(\omega)$ and imaginary~$\varepsilon_{2}(\omega)$ parts of the dielectric function~$\varepsilon(\omega)$, refractive index~$n(\omega)$, and reflectivity~$R(\omega)$ are calculated up to $15$~eV along a specified growth direction. In this work, we used~1000~$k$-points and took 0.03~eV for broadening.  
The real and imaginary dielectric functions for SL~(1-1), SL~(2-2) and SL~(3-3) are shown in figure~\ref{figure5}. 

From the quantitative point of view, the real and imaginary parts of the dielectric function are similar for all configurations. If we look at the real $\varepsilon_{1}(\omega)$ and imaginary $\varepsilon_{2}(\omega)$ parts of the dielectric function~$\varepsilon(\omega)$, we can clearly see that the passage of the real part of the dielectric function by zero is given as follows: 5.05454, 4.92727 and 5.01818~eV for SL~(1-1), SL~(2-2) and SL~(3-3), respectively. The minimum of these spectra is the energy 7.65455, 7.38182 and 7.49091~eV for SL~(1-1), SL~(2-2) and SL~(3-3), respectively. The curves of the real part at low energies show a decrease between 2.99994 and 7.57~eV for all configurations until $\varepsilon_{1}$ becomes negative. Then, the curves slowly increase to zero at higher energies, where the negative energy value of $\varepsilon_{1}(\omega)$ demonstrates that the incident photons are completely reflected. The static dielectric constants (at the limit of zero frequency) obtained in this study are as follows: 4.38, 4.52, and 4.67 for SL~(1-1), SL~(2-2) and SL~(3-3), respectively. We now turn to the analysis of the imaginary part of the dielectric function; $\varepsilon_{2}$ shows that all configurations are similar. The first peak of $\varepsilon_{2}(\omega)$ spectra is related to the optical band gap which is a threshold for optical transition between the valence band maximum and the conduction band minimum. In addition, the highest peaks in the $\varepsilon_{2}(\omega)$ of all configurations are located at 4.87496~eV.

\begin{figure}[!htb]
	\centerline{\includegraphics[scale=0.45]{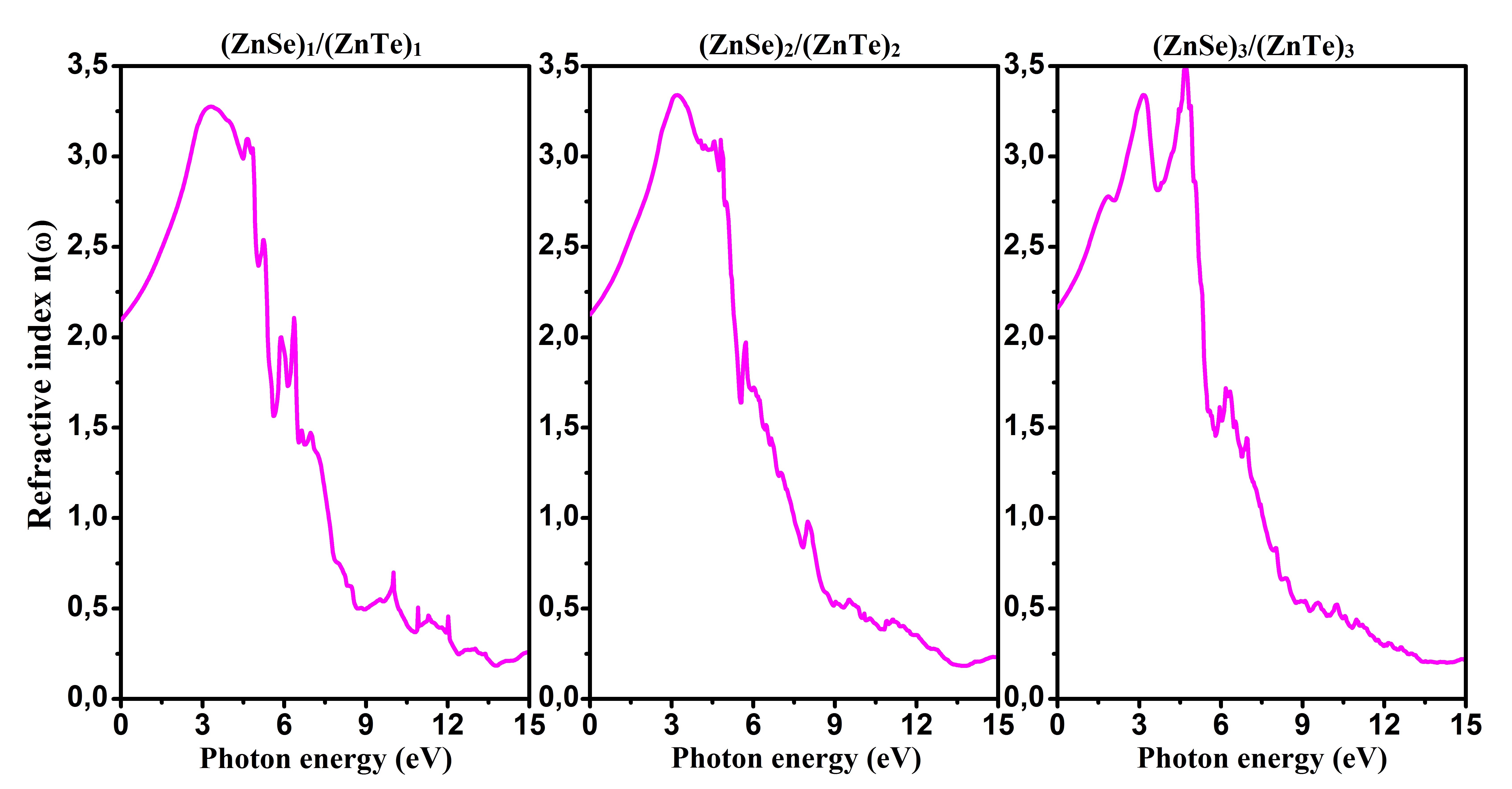}}
	\caption{(Colour online) The calculated refractive index $n(\omega)$ for (ZnSe)$_n$/(ZnTe)$_n$ superlattices.} \label{figure6}
\end{figure}

\begin{figure}[!htb]
	\centerline{\includegraphics[scale=0.40]{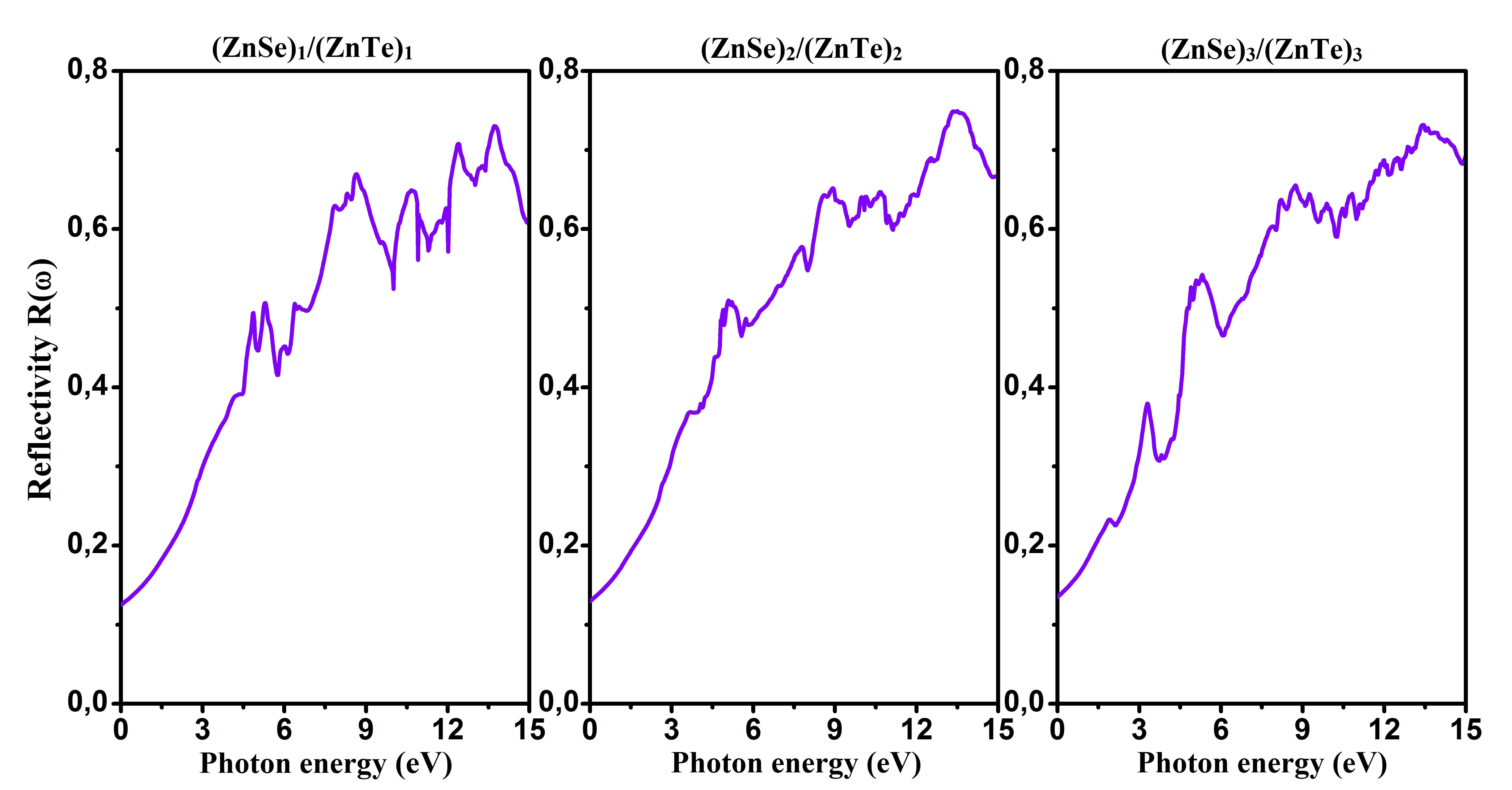}}
	\caption{(Colour online) The calculated reflectivity $R(\omega)$ for (ZnSe)$_n$/(ZnTe)$_n$ superlattices.} \label{figure7}
\end{figure}

 In figure~\ref{figure6} and figure~\ref{figure7}, the refractive index $n(\omega)$ and the reflectivity spectrum $R(\omega)$ are plotted for several different SLs~($n\text{-} n$) [SL~(1-1), SL~(2-2) and SL~(3-3)]. The optical spectra of the SLs~($n\text{-} n$) are similar. The static refractive index $n(0)$ values are observed to be 2.09, 2.013, and 2.16 for SL~(1-1), SL~(2-2) and SL~(3-3), respectively. From our examination of the reflectivity spectra of the SLs~($n\text{-} n$), the static reflectivities $R(0)$ of 12.51\%, 12.98\%, and 13.50\% are obtained in SL~(1-1), SL~(2-2) and SL~(3-3), respectively. In addition, we note that $R(\omega)$ increases up to 70\% and then starts to decrease at approximately 13.5~eV, which indicates that the SLs~($n\text{-} n$) reflects a photon, whose energy lies in the ultraviolet (UV) region. From these results, we can say that these SLs~($n\text{-} n$) behave like semiconductor compounds. As far as we know, there are no theoretical or experimental studies of the optical properties of SLs~($n\text{-} n$), so this work may be a predictive study.

\section{Conclusion}

In the present work, the FP-LMTO method is used to calculate the structural, electronic, and optical properties of (ZnSe)$_n$/(ZnTe)$_n$~SLs ($n$-$n$: 1-1, 2-2 and 3-3) based on DFT using LmtART~7.0~code. For both compounds, ZnSe and ZnTe, the calculated ground-state properties such as the lattice parameters and bulk modulus are in agreement with the available data. In comparison with the experimental data, we find that the lattice parameters are underestimated, due to the use of LDA. On the other hand, the lattice constant~$a$~({\AA}) [5.84, 11.69, and 17.55~{\AA}, for SL~(1-1), SL~(2-2) and SL~(3-3), respectively], and the derivative of bulk modulus [4.74, 4.26 and 3.74 for SL~(1-1), SL~(2-2) and SL~(3-3), respectively] at the equilibrium of (ZnSe)$_n$/(ZnTe)$_n$~SLs increase with an increase of the number of layer~$n$. Furthermore, our results for the BS and DOS, show that the ZnSe, ZnTe and (ZnSe)$_n$/(ZnTe)$_n$~SLs demonstrate a semiconductor behaviour. The bandgap $E_g$ of SLs [1.116, 1.113, and 1.059~eV for SL~(1-1), SL~($2\text{-}2$) and SL~(3-3), respectively] are direct ($\Gamma$v-$\Gamma$c) and decrease with increasing the number of layer~$n$. The optical properties  $\varepsilon_{1}(\omega)$, $\varepsilon_{2}(\omega)$, $n(\omega)$, and $R(\omega)$ are discussed up to 15~eV. The values of the static dielectric constant $\varepsilon$${}_{1}(0)$ [4.38, 4.52, and 4.67 for SL~(1-1), SL~(2-2) and SL~(3-3), respectively], the static refractive index $n(0)$ [2.09, 2.013, and 2.16 for SL~(1-1), SL~(2-2) and SL~(3-3), respectively], and the static reflectivities $R(0)$ [12.51\%, 12.98\%, and 13.50\% for SL~(1-1), SL~(2-2) and SL~(3-3), respectively] increase with increasing the number of layers $n$. The highest peaks of the reflectivity coefficient~$R(\omega)$ are detected in the UV region. The obtained results predict that the (ZnSe)$_n$/(ZnTe)$_n$ SLs are a potential candidate for quantum well lasers and solar cell heterostructures.

\ukrainianpart

\title{Електронна структура короткоперіодичних суперґраток ZnSe/ZnTe на основі розрахунків в рамках методу функціоналу густини}
\author[М. Каїд \emph{et al.}]{M.~Каїд\orcid{0000-0001-9773-8877}\refaddr{label1,label2}, 
	І.~Рашед\refaddr{label3},  Д.~Рашед\orcid{0000-0003-4686-5686}\refaddr{label2}, O.~Шереф\orcid{0000-0002-3672-1083}\refaddr{label2},  Н.~Рашед\orcid{0000-0003-3867-576X}\refaddr{label2,label4},  С.~Беналья\orcid{0000-0003-0326-8056}\refaddr{label2,label3},  M.~Mарабе\refaddr{label2,label3}} 
\addresses{
	\addr{label1} Вища нормальна школа  Бу-Саада, Факультет точних наук, Бу-Саада (28001), Мсила, Алжир
	\addr{label2} Лабораторія магнітних матеріалів, факультет точних наук, Університет Джіллалі Ліабеса, Сіді Бель-Аббес (22000), Алжир
	\addr{label3} Кафедра матеріалознавства, Факультет науки і техніки, Університет Ахмеда Бена
	Яхія Ель-Ванчарісі Тіссемсілт, Тіссемсілт (38000), Алжир
	\addr{label4} Університет Хассіби Бенбуалі в Шлефі, Факультет точних наук та інформатики, Фізичний факультет, Шлеф (02000), Алжир
}

\makeukrtitle

\begin{abstract}
	У цій роботі обговорюється вплив зміни кількості моношарів  $n$ на електронні та оптичні властивості супер\-ґраток (ZnSe)$_n$/(ZnTe)$_n$. Повні енергії розраховано за допомогою повнопотенціального методу лінійних орбіталей, для обмінно-кореляційної енергії застосовано наближення локальної густини. По-перше, розрахунки свідчать про зменшення похідної коефіцієнта об'ємної стисливості та ширини забороненої зони зі збільшенням кількості моношарів $n$. По-друге, досліджуються енергії випромінювання до $15$~еВ, діелек\-трична функція $\varepsilon(\omega$), показник заломлення $n(\omega)$ та коефіцієнт відбивання $R(\omega)$. Такі розрахунки можуть покращити розуміння властивостей короткоперіодних суперґраток (ZnSe)$_n$/(ZnTe)$_n$.
	\keywords електронна структура, оптичні властивості, суперґратки
\end{abstract}

\lastpage
\end{document}